\documentstyle[preprint,aps,epsfig]{revtex}

	{\arrayclosep 0.14em\begin{eqnarray}}{\end{eqnarray}}
\newfont{\Fr}{eufm10}   
\def\be{\begin{equation}}
\def\ee{\end{equation}}
\def\bea{\begin{eqnarray}}
\def\eea{\end{eqnarray}}
\def\b{\begin{eqnarray*}}
\def\e{\end{eqnarray*}}

\def \( {\left(}
\def \) {\right)}

\def\[{\left[}
\def\]{\right]}

\def\Journal#1#2#3#4{{#1}{\bf #2}, #4 (#3)}

\def\MEETtmp#1#2#3#4{{#1} {\it #2}, {#3}, {(#4)}}
\def\NPB{{ Nucl. Phys.} \bf B}
\def\NPBPS{{ Nucl. Phys.} B (Proc. Suppl.) }
\def\PLB{{ Phys. Lett.}  B }

\def\PRD{{ Phys. Rev.} D }
\def\PRC{{ Phys. Rep.} \bf C}

\def\PTPS{{ Prog. Theor. Phys. Suppl. }}

\begin{document}
\draft
\title{Dual Wilson Loop and Infrared Monopole Condensation
 in Lattice QCD
in the Maximally Abelian Gauge}
\author{Atsunori Tanaka and Hideo Suganuma}
\address{Research Center for Nuclear Physics (RCNP), Osaka University\\
Mihogaoka 10-1, Ibaraki, Osaka 567-0047, Japan \\
E-mail: atanaka@rcnp.osaka-u.ac.jp}
\maketitle
\begin{abstract}
Using the SU(2) lattice QCD, we formulate 
the dual Wilson loop
and
study the dual Higgs mechanism induced by monopole condensation
in the maximally abelian (MA) gauge, 
where
QCD is reduced into an abelian gauge
theory including the electric current $j_\mu$
and the monopole current $k_\mu$.
After the abelian projection in the MA gauge,
the system can be separated into 
the photon part and the monopole part
corresponding to the separation of $j_\mu$
and $k_\mu$, respectively.
We study here 
the monopole part (the monopole-current system), 
which is responsible to the electric confinement.
Owing to the absence of electric currents,
the  monopole part
is  naturally described 
using the dual gluon field $B_\mu$
without the Dirac-string singularity.
Defining the dual Wilson loop from the dual gluon $B_\mu$,
we find the perimeter law of the dual Wilson loop
in the lattice QCD simulation.
In the monopole part in the MA gauge,
the inter-monopole potential is found to be 
flat, and can be fitted as
the Yukawa potential in the infrared region
after the subtraction of the artificial
finite-size effect on the dual Wilson loop.
From more detailed analysis of 
the inter-monopole potential
considering the monopole size,
we estimate the effective dual-gluon mass $m_B \simeq 0.5$GeV 
 and the effective monopole size $R \simeq 0.2$fm.
The effective mass of the dual gluon field 
at the long distance
can be regarded as an evidence of
``infrared monopole condensation''.
\end{abstract}
\vspace{0.5cm}

\vfill\eject
\section{Introduction}
\label{section:intro}
Quantum chromo-dynamics (QCD) is 
the fundamental theory of the strong interaction
\cite{Pokorski,Cheng,Itzykson,Huang,Aitchison,Greiner,Rothe}
and is an SU($N_C$) nonabelian gauge theory
described by the quark field $q$ and the gluon field $A_\mu$ as
\begin{equation}
{\cal L}_{QCD}=-\frac{1}{2}{\rm tr}(G_{\mu\nu}G^{\mu\nu})
+\bar{q}(i/\!\!\!\!D - m_q)q,
\end{equation}
where $G_{\mu\nu}$ is the SU($N_c$) field strength
$G_{\mu\nu} \equiv \frac{1}{ie}[{D}_\mu, {D}_\nu]$
with the covariant-derivative 
operator ${D}_\mu={\partial}_\mu 
+ ieA_\mu$.
Due to the asymptotic freedom, which
 is one of the most important features in QCD
the gauge-coupling constant of
QCD becomes small in the ultraviolet region 
\cite{Pokorski,Cheng,Itzykson}.
Accordingly, the perturbative QCD can describe
 the high-energy phenomena like 
the Bjorken scaling and the hadron jet properties
\cite{Huang,Aitchison,Greiner}.

On the other hand, in the low-energy region,
the QCD-coupling constant becomes large,
and there arise the nonperturbative-QCD (NP-QCD)
phenomena such as color confinement
and dynamical chiral-symmetry breaking
corresponding to the strong-coupling nature.
These NP-QCD phenomena
are extremely difficult to understand 
in the analytical manner
from QCD, and
have been studied by using the effective models
\cite{Greiner}
or the lattice QCD simulation \cite{Rothe}. 
Here, the lattice QCD Monte Carlo simulation is the numerical
calculation of the QCD partition functional,
and it is one of the most reliable methods directly
based on QCD.
In fact, the lattice QCD simulations well reproduce
nonperturbative quantities
such as the quark static potential,
the chiral condensate
$\langle {\bar q}q\rangle$
and low-lying hadron masses \cite{Rothe}.

Recently, the lattice QCD simulation
has shed light on the confinement mechanism
in terms of the dual-superconductor
picture,
which was proposed by
Nambu,'t Hooft and Mandelstam 
in the middle of 1970's
\cite{Nambu,tHooft75,Mandelstam}.
In this scenario,
quark confinement can be understood
with the dual version of the superconductivity.
In the ordinary superconductor, the Meissner effect
occurs by condensation 
of the Cooper-pair with the electric charge.
Consider the existence of the magnetic charges with the 
opposite sign immersed in the superconductor,
then the magnetic flux is squeezed like a tube 
between the magnetic charges, and 
the magnetic potential between them
becomes linear as the result of the Meissner effect
\cite{Abrikosov}.
In the dual-superconductor scenario, 
the QCD vacuum is assumed as the dual version of the superconductor,
and the dual Meissner effect
brings the one-dimensional flux squeezing \cite{Haymaker}
between the quark and the anti-quark,
which leads to the linear confinement potential 
\cite{Huang,Aitchison,Greiner,Rothe,Nambu,SST}.

The dual Higgs mechanism, however, requires 
``color-magnetic monopole condensation'' as 
the dual version of electric condensation
in the superconductor,
although QCD dose not include 
{\rm the color-magnetic monopole} 
as the elementary degrees of freedom.
On the appearance of magnetic monopoles from QCD,
't Hooft showed that
QCD is reduced to an abelian gauge theory with magnetic monopoles 
by taking the abelian gauge,
which fixes the partial gauge symmetry SU($N_C$)/U$(1)^{N_c-1}$
through the diagonalization of 
a gauge-dependent variable \cite{tHooft81}.
Here,
the monopole appears as the topological object
corresponding to the nontrivial homotopy group
$\pi_{2}$(SU($ N_{c}$)/U$(1)^{N_{c}-1}$)=$Z_{\infty}^{N_{c}-1}$.

As for the irrelevance of off-diagonal gluons,
recent lattice QCD studies show the abelian dominance \cite{Ezawa}
for the NP-QCD phenomena in the maximally abelian (MA) gauge
\cite{Kronfeld,Brandstater}.
For instance,
confinement \cite{Yotsuyanagi,Hioki91}
and 
dynamical chiral-symmetry breaking \cite{Miyamura,Woloshyn}
are almost described only by the diagonal gluon component,
in the MA gauge.
Then, taking the MA gauge and removing
off-diagonal gluons, the abelian-projected QCD (AP-QCD) is
obtained as the abelian gauge theory keeping the NP-QCD features.
AP-QCD includes not only the electric 
current $j_\mu$ but also the magnetic current $k_\mu$,
and can be decomposed into the monopole part and the photon part
corresponding to the separation of $k_\mu$ and $j_\mu$,
respectively \cite{DeGrand,Stack_Wensley}. 
The lattice QCD studies show that only 
the monopole part
is responsible to
NP-QCD phenomena \cite{Miyamura,Stack_Wensley,Bali,Polikarpov,STSM}
especially to the electric confinement 
\cite{Stack_Wensley,Bali,Polikarpov}
in the MA gauge.
This is called as the monopole dominance.

In this paper, we concentrate the monopole part (the monopole-current system)
in the MA gauge in QCD,
and study the dual Higgs mechanism in the QCD vacuum
based on the dual gauge formalism \cite{INNOCOM_Tanaka,YKIS_Suganuma}.
To this end,
we perform the SU(2) lattice QCD simulation in the MA gauge,
and extract the monopole current $k_\mu$ 
as the relevant degrees of freedom for electric confinement.
Then, we
calculate inter-monopole potential in the monopole part,
(the monopole current system) in QCD
to examine monopole condensation,
and evaluate
the effective mass of the dual gluon field $B_\mu$
\cite{INNOCOM_Tanaka,YKIS_Suganuma}. 
\section{Separation of AP-QCD into the Monopole Part and the Photon Part in the MA Gauge}
\subsection{MA Gauge Fixing and AP-QCD}
Recent studies with
the lattice QCD Monte Carlo simulation
have revealed the abelian dominance and the monopole dominance
in the maximally abelian (MA) gauge
for the nonperturbative QCD (NP-QCD) phenomena
such as confinement, dynamical chiral-symmetry breaking
and instantons [18-26].

In the continuum Euclidean QCD with $N_c=2$,
the MA gauge fixing
is defined by the minimizing
the total amount of off-diagonal gluons,
\begin{equation}
R^{\rm }_{\rm off}= 
\int d^{4}x
[\{A_{1}^{\mu}(x)\}^{2}+\{A_{2}^{\mu}(x)\}^{2}]=
2\int d^{4}x [A_{+}^{\mu}(x)A_{-}^{\mu}(x)]
\label{eq:MA-gaugesu2}
\end{equation}
with
\begin{eqnarray}
A^{\rm}_{\mu}(x)&=&\sum_{a=1}^{3}A^{a}_{\mu}(x)\frac{{\tau}^{a}}{2},
\\
A^{\pm}_\mu(x)&\equiv& \frac{1}{\sqrt{2}}
\{A^1_\mu (x)\pm i A^2_\mu (x)\},
\end{eqnarray}
by the SU(2)-gauge transformation.
In the MA gauge, off-diagonal gluon components, 
$A^1_{\mu}$ and $A^2_{\mu}$,
become as small as possible, and
 the gluon field $A^{\rm }_{\mu}$ 
mostly approaches to the abelian gluon 
field,
$A^{\rm Abel}_{\mu} \equiv 
A^{3}_{\mu}\frac{\tau^{3}}{2}$.
In the MA gauge, the SU(2) local
symmetry is reduced into the U$(1)_3$ local symmetry with the global Weyl 
symmetry.
Under the residual $U(1)_3$-gauge transformation
with $\omega = \exp(-i\phi\frac{\tau^3}{2}) \in$ U$(1)_3$,
the gluon components are transformed as
\begin{eqnarray}
A^3_{\mu}(x) &\to&A^{3}_{\mu}(x)+\partial_\mu \phi(x),
\\
A^{\pm}_\mu(x)&\to& e^{i\phi(x)}A^{\pm}_\mu(x).
\end{eqnarray}
Then, in the MA gauge,
the diagonal gluon $A^3_\mu$ behaves as the abelian gauge field,
while off-diagonal gluons $A^\pm_\mu$ behave as charged matter fields
in terms of the residual gauge symmetry.

As a remarkable feature of the MA gauge,
the abelian dominance holds for the NP-QCD phenomena
such as quark confinement and chiral-symmetry breaking
[18-21].
Here, we call abelian dominance for an operator $\hat O$,
when the expectation value $\langle O[A_\mu] \rangle$
is almost equal to the expectation value 
$\langle O[A^{\rm Abel}_\mu] \rangle_{\rm MA}$, 
where off-diagonal gluons are dropped off 
in the MA gauge.
For instance, the abelian string tension 
$\sigma_{\rm Abel}\equiv\langle \sigma(A^{\rm Abel}_\mu) \rangle_{\rm MA}$
in the MA gauge is almost equal to 
$\sigma_{\rm SU(2)}\equiv\langle \sigma(A^{\rm SU(2)}_\mu) \rangle$
as $\sigma_{\rm Abel} \simeq 0.92\sigma_{\rm SU(2)}$ for $\beta \simeq 2.5$
in the lattice QCD 
[24,25].
Thus, NP-QCD phenomena 
are almost reproduced only by the abelian gluon $A^{\rm 
 Abel}_{\mu}$,
 and off-diagonal gluon components $A^{\pm}_\mu$
 do not contribute to NP-QCD in the MA gauge.
Hence, as long as the infrared physics is concerned,
QCD in the MA gauge
can be approximated by the abelian projected QCD (AP-QCD),
where the SU(2) gluon field $A_\mu$ is replaced by
the abelian gluon field $A^{\rm Abel}_\mu$.
In other word, AP-QCD is the abelian gauge theory
keeping essence of NP-QCD, and is extracted from
QCD in the MA gauge.
Hereafter, we pay attention to the AP-QCD described
by $A^{\rm Abel}_\mu$ in the MA gauge.

The abelian-projected QCD (AP-QCD)
includes not only $j_\mu$ 
but also $k_\mu$.
In this subsection, we investigate the general argument
on the extended electro-magnetic system
including both
 the electric current $j_\mu$ and the magnetic current $k_\mu$.
In the extended Maxwell equations with $j_\mu$ and $k_\mu$,
the field strength $F_{\mu\nu}^{\rm Abel}$ satisfies as 
\begin{eqnarray}
\partial_{\mu}F^{\rm Abel}_{\mu\nu}&=&j_\nu\\
\partial_{\mu}{~^{*}\!F^{\rm Abel}_{\mu\nu}}
=-\partial_{\mu}{~^{*}\!\xi_{\mu\nu}}&=&k_\nu
\end{eqnarray}
with $^*\!F^{\rm Abel}_{\mu\nu}
\equiv \frac12 \varepsilon_{\mu\nu\rho\sigma}
F^{\rm Abel}_{\rho\sigma}$.
In the presence of both $j_\mu$ and $k_\mu$,
the field strength $F^{\rm Abel}_{\mu\nu}$ 
cannot be described by the simple two-form
$\partial_\mu A_\nu^{\rm Abel} - \partial_\nu A_\mu^{\rm Abel}$
 with the regular one-form $A^{\rm Abel}_\mu$ [29].
In general, the field strength $F^{\rm Abel}_{\mu\nu}$
consists of two parts,
\begin{equation}
F_{\mu\nu}^{\rm Abel}=
(\partial \land A^{\rm Abel})_{\mu\nu}
+~^*\!\xi_{\mu\nu},
\label{eq:twoform}
\end{equation}
where the former part denotes the ordinary two-form and 
the latter part
$^*\!\xi_{\mu\nu}=\frac{1}{2}\epsilon_{\mu\nu\alpha\beta}
\xi_{\alpha\beta}$
denotes the Dirac-string singularity [29].
Here, $\xi_{\mu\nu}$ can be written as
\begin{eqnarray}
\xi_{\mu\nu}(x)&=&\frac{1}{n \cdot \partial}(n \land k)_{\mu\nu}
\nonumber\\
&=& \int d^4y \langle x|\frac{1}{n \cdot \partial} |y \rangle
(n_\mu k_\nu(y)-n_\nu k_\mu(y))
\nonumber\\
&=& \int d^4y \theta(x_n-y_n)\delta^3({\bf x}^\bot-{\bf y}^\bot) 
(n_\mu k_\nu(y)-n_\nu k_\mu(y)),
\end{eqnarray}
where $x_n \equiv x_\mu n_\mu$ and 
$x^\bot_\mu \equiv x_\mu -(x \cdot n) n_\mu$.
%
Here, $n_\mu$ is arbitrary four-dimensional unit vector
corresponding to the direction of the Dirac string.
Thus, in the ordinary description 
[29]
the system includes the 
singularity as the Dirac string $\xi_{\mu\nu}$,
which makes the analysis complicated.

\subsection{Separation of AP-QCD into Photon Part and Monopole Part}

To clarify the roles of $j_\mu$ and $k_\mu$
to the nonperturbative quantities of QCD,
we consider the decomposition of AP-QCD
into the photon part
and the monopole part,
corresponding to the separation of $j_\mu$ and $k_\mu$.
We call this separation into the photon and monopole parts
as the ``photon projection'' and the ``monopole projection'', respectively.
[26,30].

The field strength $F^{\rm Abel}_{\mu\nu}$
in AP-QCD is separated into $F^{\rm Ph}_{\mu\nu}$ in the photon part 
and 
$F^{\rm Mo}_{\mu\nu}$ in the monopole part
\begin{eqnarray}
F_{\mu\nu}^{\rm Abel}=F_{\mu\nu}^{\rm Ph}+F_{\mu\nu}^{\rm Mo}.
\label{eq:F=F+F}
\end{eqnarray}
As the physical requirement, $F^{\rm Ph}_{\mu\nu}$ and $F^{\rm Mo}_{\mu\nu}$
satisfies the Maxwell equations as
\begin{eqnarray}
\partial_\mu F_{\mu\nu}^{\rm Ph}&=&j_\nu,~~
\partial_\mu ~^*\!F_{\mu\nu}^{\rm Ph}=0,
\label{eq:M-eq-ph}\\
\partial_\mu F_{\mu\nu}^{\rm Mo}&=&0,~~
\partial_\mu ~^*\!F_{\mu\nu}^{\rm Mo}=k_\nu,
\label{eq:M-eq-mo}
\end{eqnarray}
respectively.
From the requirement $(\ref{eq:M-eq-ph})$,
there exists the regular vector $A^{\rm Ph}_\mu$ 
is defined so as to satisfying
\begin{eqnarray}
F^{\rm Ph}_{\mu\nu}
&=&
\partial_\mu A_\nu^{\rm Ph} - \partial_\nu A_\mu^{\rm Ph}
\label{eq:def-ph}
\end{eqnarray}
in the photon part.
This relation is the same as
$F_{\mu\nu}=\partial_\mu A_\nu^{\rm} - \partial_\nu A_\mu^{\rm}$
in the ordinary QED.

As for the monopole part, 
$F^{\rm Mo}_{\mu\nu}$ cannot be expressed by the simple two-form,
but is expressed as
\begin{eqnarray}
\partial_\mu A_\nu^{\rm Mo} - \partial_\nu A_\mu^{\rm Mo}
&=&F^{\rm Mo}_{\mu\nu}+~^*\!\xi_{\mu\nu},
\label{eq:def-mo}
\end{eqnarray}
where the second term provides the breaking of the Bianchi identity.
From Eqs.(\ref{eq:twoform}), (\ref{eq:F=F+F}), (\ref{eq:def-ph})
and (\ref{eq:def-mo}),
one finds the relation
satisfy
\begin{eqnarray}
\partial \land A^{\rm Abel} = \partial \land (A^{\rm Ph}+A^{\rm Mo}),
\label{eq:two form eq}
\end{eqnarray}
and then the difference between $A_\mu^{\rm Abel}$ and 
$A_\mu^{\rm Ph}$+$A_\mu^{\rm Mo}$ is at most the total differential term
as $\partial_\mu \chi$.
Hence, 
we can set 
\begin{eqnarray}
A_\nu^{\rm Abel} = A_\nu^{\rm Ph}+A_\nu^{\rm Mo}
\label{eq:two form eq}
\end{eqnarray}
by taking a suitable gauge
without loss of generality.

In the practical calculation,
the photon part variable
$A_\mu^{\rm Ph}$ 
and the monopole part variable $A_\mu^{\rm Mo}$ can be obtained as
\begin{eqnarray}
A^{\rm Ph}_{\nu} &=& \Box^{-1}j_{\nu}
=\Box^{-1}\partial_{\mu}F^{\rm Abel}_{\mu\nu},
\label{eq:ph}\\
A^{\rm Mo}_{\nu} &=& 
\Box^{-1}\partial_{\mu}~^*\!\xi_{\mu\nu},
\label{eq:mo}
\end{eqnarray}
where the inverse d'Alembertian 
$\Box^{-1}$ is the nonlocal operator 
[31],
\begin{eqnarray}
\langle x|\Box^{-1}|y \rangle
= -\frac{1}{4 \pi^2}\frac{1}{(x-y)^2},
\end{eqnarray}
which satisfies 
$\Box_x \langle x|\Box^{-1}| y \rangle=\langle x|y \rangle=\delta^4(x-y)$.

Let us check that $A^{\rm Ph}_\mu$ and $A^{\rm Mo}_\mu$
defined in Eqs.(\ref{eq:ph}) and (\ref{eq:mo}),
satisfies the physical requirement in Eqs.(\ref{eq:def-ph}) and 
(\ref{eq:def-mo})
First, we consider the photon part with $A^{\rm Ph}_\mu$.
Starting from Eq.(\ref{eq:ph}),
one easily finds $\partial_\mu A^{\rm Ph}_\mu =0$,
and hence the first Maxwell equation in Eq.(\ref{eq:M-eq-ph}) in the photon 
part can be derived as 
\begin{eqnarray}
\partial_{\mu}F_{\mu\nu}^{\rm Ph}&=&
\partial_{\mu}(\partial_{\mu}A^{\rm Ph}_\nu- \partial_\nu A^{\rm Ph}_\mu)
=\Box A^{\rm Ph}_\nu - \partial_\nu (\partial_\mu A^{\rm Ph}_\mu)
=\Box A^{\rm Ph}_\nu
\nonumber \\
&=& \Box \Box^{-1}j_{\nu}=j_\nu.
\
\end{eqnarray}
The second Maxwell equation 
in Eq.(\ref{eq:M-eq-ph}) is automatically derived as  
the Bianchi identity for the two form of $A_\mu^{\rm Ph}$.
Thus, the photon part with $A_\mu^{\rm Ph}$ defined by Eq.(\ref{eq:ph}) 
does not include the magnetic current
but only includes the electric current.

Second, we consider the monopole part with $A^{\rm Mo}_\mu$.
From Eqs.(\ref{eq:twoform}), (\ref{eq:ph}) and (\ref{eq:mo}),
 the sum of $A^{\rm Ph}_\mu$ and $A^{\rm Mo}_\mu$
can be written as
\begin{eqnarray}
A^{\rm Ph}_\mu+A^{\rm Mo}_\mu 
&=&
\Box^{-1}\partial_{\alpha}
(F^{\rm Abel}_{\alpha\mu}+~^*\!\xi_{\alpha\mu})
=\Box^{-1}\partial_{\alpha}
(\partial_\alpha A_\mu^{\rm Abel} - \partial_\mu A_\alpha^{\rm Abel})
\nonumber \\
&=&A^{\rm Abel}_\mu-\partial_\mu (\partial_\alpha A_\alpha^{\rm Abel}),
\end{eqnarray}
and hence the two-form of $A^{\rm Mo}_\mu$ satisfies the physical requirement 
$(\ref{eq:def-mo})$ as
\begin{eqnarray}
(\partial_\mu A_\nu^{\rm Mo} - \partial_\nu A_\mu^{\rm Mo})&=&
\partial_\mu A_\nu^{\rm Abel} - \partial_\nu A_\mu^{\rm Abel}
-(\partial_\mu A_\nu^{\rm Ph} - \partial_\nu A_\mu^{\rm Ph})
\nonumber \\
&=&F^{\rm Abel}_{\mu\nu}+^*\!\xi_{\mu\nu}-F^{\rm Ph}_{\mu\nu}
\nonumber \\
&=&F^{\rm Mo}_{\mu\nu}+^*\!\xi_{\mu\nu}. 
\end{eqnarray}
Therefore, in the monopole part, 
the Maxwell equations
is derived as
\begin{eqnarray}
\partial_{\mu}F_{\mu\nu}^{\rm Mo}=
\partial_{\mu}F_{\mu\nu}^{\rm Abel}-\partial_{\mu}F_{\mu\nu}^{\rm Ph}
=0,\\
\partial_{\mu}~^*\!F_{\mu\nu}^{\rm Mo}=
\partial_{\mu}~^*\!F_{\mu\nu}^{\rm Abel}
-\partial_{\mu}~^*\!F_{\mu\nu}^{\rm Ph}=
k_\nu,
\end{eqnarray}
starting from Eq.(\ref{eq:mo}).

Thus, $A^{\rm Ph}_\mu$ and $A^{\rm Mo}_\mu$ 
defined as Eqs.(\ref{eq:ph}), and (\ref{eq:mo})
satisfy the physical requirement Eqs.(\ref{eq:def-ph})
and (\ref{eq:def-mo}).
Hence, the monopole part carries the same amount of the magnetic
current as that is the original abelian sector,
whereas it dose not 
carries the electric current.
The situation is just the opposite in the photon part. 
In the actual lattice QCD simulation,
the monopole current $k_\mu$ and 
the electric current $j_\mu$ are slightly modified through
the monopole  and the photon projections, respectively,
due to the numerical error on the lattice.
However, these differences are negligibly small in the actual
lattice QCD simulation.
In fact, $k_\mu^{\rm Mo} \simeq k_\mu$ and $j_\mu^{\rm Mo} \simeq 0$
hold in the monopole part, and $j_\mu^{\rm Ph} \simeq j_\mu$
and $k_\mu^{\rm Ph} \simeq 0$ hold in the photon part within 1$\%$ error. 
Here, we have kept the labels as ``$Mo$'' and ``$Ph$'' 
for the electric current and the monopole current,
and we have used $(k_{\mu}^{\rm Mo},j_{\mu}^{\rm Mo})$
and $(k_{\mu}^{\rm Ph},j_{\mu}^{\rm Ph})$ for these currents
in the monopole part and the photon part, respectively.

As a remarkable fact,
lattice QCD simulations show that nonperturbative 
quantities such as the string tension,
the chiral condensate and instantons 
 are almost reproduced only by the monopole part
in the MA gauge, which is called as monopole dominance 
[23-26]. 
On the other hand,
the photon part dose not contribute 
these nonperturbative quantities in QCD.

Since we are interested in the NP-QCD phenomena,
it is convenient and transparent to
extract the relevant degrees of freedom for NP-QCD
by removing irrelevant degrees of freedom like 
the off-diagonal gluons $A^{\pm}_\mu$
and the electric current $j_\mu$.
Therefore, we
concentrate ourselves to the monopole part,
which keeps the essence of NP-QCD as confinement.

\section{Dual Gauge Formalism\\
-Dual Gluon Field and Dual Wilson Loop-}

In this section, we study the monopole part
of the QCD vacuum 
using the dual gauge formalism 
[27,28].
In the MA gauge, 
the monopole part carries essence of the 
nonperturbative QCD as the electric confinement.
According to the absence of the electric current $(j_\mu=0)$,
the Maxwell equation in the monopole part 
becomes
\begin{eqnarray}
\partial_{\mu}F^{\rm Mo}_{\mu\nu}&=&0\\
\partial_{\mu}{^{*}\!F^{\rm Mo}_{\mu\nu}}&=&k_{\nu},
\end{eqnarray}
where $F^{\rm Mo}_{\mu\nu}$ denotes the field strength in the monopole part.
This system resembles the dual version of QED
with $j_\mu \ne 0$ and $k_\mu=0$, and hence
it is useful to introduce 
the dual gluon field $B_{\mu}$ in the monopole part 
for the study of the dual Higgs mechanism in QCD
[27,28].

The dual gluon field $B_{\mu}$ is defined so as to satisfy 
the relation 
\begin{equation}
\partial_\mu B_\nu - \partial_\nu B_\mu=^*\!F^{\rm Mo}_{\mu\nu},
\label{eq:def-bmu}
\end{equation}
which is the dual version of the ordinary relation
$F_{\mu\nu} \equiv \partial_\mu A_\nu^{\rm Mo} - \partial_\nu A_\mu^{\rm Mo}$
in QED. 
The interchange between $A_\mu$ and $B_\mu$
corresponds to the electro-magnetic duality transformation,
$F_{\mu\nu} \leftrightarrow ~^*\!F_{\mu\nu}$ or
${\bf H} \leftrightarrow {\bf E}$.
Owing to the absence of $j_\mu$, 
the dual gauge field $B_\mu$
can be introduced without the Dirac-string singularity. 
In the other words,
the absence of $j_\nu$ is automatically derived as
the dual Bianchi identity, 
\begin{equation}
j_\nu=\partial_\mu F_{\mu\nu}
=\partial_{\mu}~^*\!(\partial \land B)_{\mu\nu}=0.
\label{eq:bianchi}
\end{equation}

Let us consider the derivation of the 
dual gauge field $B_\mu$ from the monopole current $k_\mu$.
Taking the dual Landau gauge $\partial_{\mu}B_\mu=0$,
the Maxwell equation
$\partial_\mu~^*\!F^{\rm Mo}_{\mu\nu}
=\partial^2 B_\nu - \partial_\nu(\partial_\mu B_\mu)
=k_\nu$ is simply reduced as $\Box B_\mu=k_\mu$.
Therefore,
the dual gluon field $B_\mu$ 
is obtained
by using the inverse d'Alembertian $\Box^{-1}$
as
\begin{eqnarray}
B_{\nu}(x)&=&\Box^{-1}k_{\nu}
\end{eqnarray}
or equivalently
\begin{eqnarray}
B_{\nu}(x)&=&\int d^4y \langle x|\Box^{-1}|y \rangle k_\nu(y)
= -\frac{1}{4 \pi^2}\int d^4y \frac{1}{(x-y)^2}k_\nu(y).
\end{eqnarray}
Thus, the monopole part is described 
by the monopole current $k_\mu$
 and the dual gluon $B_\mu$ in the regular manner
based on the dual gauge formalism.

In the dual superconductor picture in QCD,
$k_\mu$ and $B_\mu$ correspond to
the Cooper-pair and the photon 
in the superconductor, respectively.
The Cooper-pair and the photon are essential
degrees of freedom which bring the superconductivity.
In the superconductor,
the photon field $A_\mu$
gets the effective mass
as the result of Cooper-pair condensation,
and this leads to the Meissner effect. 
Accordingly, the potential between the static electric charges
becomes the Yukawa potential $V_Y(r) \propto \frac{e^{-mr}}{r}$
in the ideal superconductor obeying the London equation.
Similarly,
the dual gluon $B_\mu$ is expected to be massive
in the  the monopole-condensed system,
and the mass acquirement of $B_\mu$
leads to the dual Meissner effect.
In other words,
the acquirement of dual gluon mass $m_B$
reflects monopole condensation,
and brings electric confinement.
Hence, we can investigate the dual Higgs mechanism in QCD
by evaluating the dual gluon mass $m_B$,
which is estimated from the inter-monopole potential.

%
To estimate the interaction between the monopoles, 
we propose the dual Wilson loop $W_D$ 
[27,28].
The dual Wilson loop $W_D$ is defined 
by the line-integral of the dual gluon field 
$B_\mu \equiv B_\mu^3\frac{\tau^3}{2}$
along a closed loop $C$,
\begin{equation}
W_D(C) \equiv
\frac{1}{2}{\rm tr}\exp{(ie_{M}\oint_C B_\mu dx_\mu)}
=Re[\exp{(i\frac{e_{M}}{2}\oint_C B_\mu^{3}dx_\mu)}],
\label{eq:def-dw}
\end{equation}
which is the {\it dual version of the abelian Wilson loop} 
\begin{equation}
W_{\rm Abel}(C)
\equiv
\frac{1}{2}{\rm tr}\exp({ie\oint_C A^{\rm Abel}_\mu dx_\mu})
=Re[\exp{(i\frac{e}{2}\oint_C A_\mu^{3}dx_\mu)}].
\label{eq:def-w2}
\end{equation}
Using the Stokes theorem, the dual Wilson loop
$W_D(C)$ is rewritten as
\begin{equation}
W_D(C) =\frac{1}{2}{\rm tr}\exp{(ie_{M}
\int_S ~^*\!F^{\rm Mo}_{\mu\nu}dS_{\mu\nu})},
\label{eq:def-dw-st}
\end{equation}
with the dual gauge field strength $F^{\rm Mo}_{\mu\nu}$.
The dual Wilson loop
$W_D(R \times T)$
 describes the interaction
between
the monopole-pair with the test magnetic charges
$\frac{e_{M}}{2}$ and $-\frac{e_{M}}{2}$.
Here, these test magnetic charges are
pair-created at $t=0$
and are pair-annihilated at $t=T$
keeping the spatial distance $R$
for $0\leq t \leq T$.
As $T$ goes to infinity, 
the dual Wilson loop 
$\langle W_D(R \times T) \rangle$
means the interaction 
between the static monopole and 
anti-monopole with
the separation of the distance $R$.
The inter-monopole potential
is obtained from the dual Wilson loop as
\begin{equation}
V_{M}(R) = -\lim_{T \to \infty} \frac{1}{T}\ln 
\langle W_D(R \times T)\rangle
\label{eq:imp}
\end{equation}
in a similar manner to the extraction of the inter-quark potential
from the Wilson loop 
[4-7].
To summarize here, for the investigation of the dual Higgs mechanism in QCD, 
we have introduced the dual gluon field $B_\mu$
and the dual Wilson loop
in the monopole part of the AP-QCD in the MA gauge.
In the next section, 
we consider the practical procedure on
the calculation of the dual Wilson loop
and the inter-monopole potential in the lattice QCD formalism.


\section{Dual Gauge Formalism on the Lattice}

We study the dual Wilson loop and
the inter-monopole potential in the MA gauge
using the SU(2) lattice QCD Monte Carlo simulation.
The lattice QCD simulation is the direct calculation 
of the QCD partition functional using the Monte Carlo method.
The physical expectation value of an observable $O[A_\mu]$
is numerically obtained by averaging its value over all gauge 
configurations with the weight factor $\exp(-S[A_\mu])$,
where  $S[A_\mu]$ denotes the lattice QCD action 
in the
Euclidean metric \cite{Rothe}.

In the lattice gauge formalism
with the lattice spacing $a$, the SU(2) link variable is defined by 
$U_\mu(s) \equiv \exp(iaeA_{\mu}(s))=\exp(iae A_\mu^a(s) \tau^{a}/2) 
~\in$ SU(2),
where $e$ and $\tau^a/2$ denote the QCD gauge coupling and 
the generator of the SU(2) group, respectively.
The standard lattice QCD action in the gauge sector 
is defined by
\begin{eqnarray}
S^{\rm L}=
\beta \sum_{s, \mu, \nu}[1-\frac{1}{2N_c}
{\rm tr}\{ U_{\mu\nu}(s)+U_{\mu\nu}^{\dagger}(s) \}], 
~~~ \beta \equiv \frac{2N_c}{e^2}
\label{eq:lattice action}
\end{eqnarray}
using the plaquette variable
$U_{\mu\nu}(s)\equiv U_\mu(s)U_\nu(s+\mu)
U_\nu^{\dagger}(s+\nu)U_\mu^{\dagger}(s)$.
In the continuum limit $a \to 0$,
the plaquette $U_{\mu\nu}(s)$ becomes 
$\exp[ia^2G_{\mu\nu}(s)]$, and hence
$S^{\rm L}$ coincides with the continuum QCD action
\begin{eqnarray}
S=\int d^4x \frac{1}{2}{\rm tr}(G_{\mu\nu}G_{\mu\nu}).
\label{eq:continuume action}
\end{eqnarray}
Thus, the QCD system is described by the link variable 
$U_\mu(s) \in$ SU($N_c$) 
instead of  the gauge field $A_\mu(x) \in su(N_c)$ in the lattice formalism.


Here, we consider the extraction of the abelian-projected QCD (AP-QCD)
from the lattice QCD.
%
In the SU(2) lattice formalism,
the MA gauge fixing is achieved by maximizing
\begin{equation}
R=
\frac{1}{2}{\rm tr}\sum_{s, \mu}
[\tau^{3}U_\mu (s){\tau^3}U^{\dagger}_\mu (s)]
=\sum_{s, \mu}\left[1-2 \left(
\{ U_\mu^{1}(s) \} ^{2}+ \{ U_\mu^{2}(s) \} ^{2}
\right) \right]
\label{eq:MAfix}
\end{equation}
by the SU(2) gauge transformation,
\begin{equation}
U_\mu \to U^{\rm MA}_\mu=V(s)U_\mu(s)V^\dagger(s+{\hat \mu}), 
\label{eq:MA-tr}
\end{equation}
where $V(s)$ and $V(s+{\hat \mu})$
are the gauge functions located at the starting 
and end points of the link variable $U_\mu(s)$.
In this gauge, 
the absolute value of off-diagonal components $U^1_\mu(s)$
and $U^1_\mu(s)$ 
are forced to be small as possible
using the gauge degrees of freedom.

In accordance with the Cartan decomposition, 
the SU(2) link variable $U_\mu(s)$ is factorized as 
\begin{eqnarray}
U^{\rm MA}_{\mu}(s)&=&M_{\mu}(s)u_{\mu}(s)~,\\
M_{\mu}&\equiv&
\exp[i(\theta_{\mu}^{1}\tau^{1}+\theta_{\mu}^{2}\tau^{2})],
~~u_{\mu}(s)\equiv \exp[i(\theta_{\mu}^{3}\tau^{3})],
\label{eq:M-def}
\end{eqnarray}
where $u_{\mu} \in {\rm U(1)_{3}}$
and $M_{\mu} \in {\rm SU(2)/U(1)_{3}}$ 
correspond to 
the diagonal part and the off-diagonal part of the gluon field,
respectively.
In the continuum limit,
the angle variable $\theta^a_\mu$ goes to the gluon field $A^a_\mu$
as 
$\theta^a_\mu \to \frac{1}{2}eaA^a_\mu$.
The off-diagonal factor $M_\mu(s)$
is rewritten as
\begin{eqnarray}
M_{\mu}(s)=
e^{i(\theta_{\mu}^{1}\tau^{1}+\theta_{\mu}^{2}\tau^{2})}
&=&
{\left(
\begin{array}{cc}
\cos \theta_{\mu}^{\rm } & -\sin \theta_{\mu}^{\rm }e^{-i\chi_{\mu}} \\
\sin \theta_{\mu}^{\rm }e^{i\chi_{\mu}} & \cos \theta_{\mu}^{\rm }
\end{array}
\right)}
\end{eqnarray}
with
\begin{equation}
\theta_{\mu}^{\rm } \equiv {\rm mod}_{\pi/2}
\sqrt{(\theta_{\mu}^{1})^{2}+(\theta_{\mu}^{2})^{2}}~~\in [0, \frac{\pi}{2}],
~~~~~\chi \equiv \tan^{-1}\frac{\theta_{\mu}^{1}}{\theta_{\mu}^{2}}.
\nonumber
\end{equation}
Under the abelian gauge transformation with $v(s) \in {\rm U(1)_{3}}$,
$M_\mu(s)$ and $u_{\mu}(s)$ are transformed as
\begin{eqnarray}
M_{\mu}(s) \to M^v_{\mu}(s)=v(s)M_{\mu}(s)v^{\dagger}(s),\\
u_{\mu}(s) \to u^v_{\mu}(s)=v(s)u_{\mu}(s)v^{\dagger}(s+{\hat \mu}),
\end{eqnarray}
to  keep the form of Eq.(\ref{eq:M-def})
for $M^v_\mu \in$ SU(2)/U$(1)_3$ and $u^v_\mu \in$ U$(1)_3$.
Then, $M_\mu(s)$ behaves as the charged matter field
and the abelian link-variable
\begin{equation} 
u_{\mu}(s)=
\left(
\begin{array}{cc}
e^{i\theta_{\mu}^{3}} & 0 \\
0 & e^{-i\theta_{\mu}^{3}} 
\end{array}
\right)
\end{equation}
behaves as a abelian gauge field
with respect to the residual abelian gauge symmetry.

In the lattice QCD, the abelian dominance is expressed as 
$\langle O(U_\mu) \rangle \simeq \langle O(u_\mu) \rangle_{\rm MA}$
for the infrared quantities such as $\sigma$
and $\langle {\bar q}q \rangle$ in the MA gauge.
The abelian projection is performed by the replacement of 
$U_\mu(s) \to u_\mu(s)$,
and then
the abelian-projected QCD
(AP-QCD) on the lattice is described only with the abelian link-variable
$u_{\mu}(s)$,
which is the diagonal part 
of the SU(2) link-variable $U^{\rm MA}_{\mu}(s)$
in the MA gauge.

Next, we consider the separation of AP-QCD into the monopole part
and the photon part in the lattice formalism.
Using the diagonal gluon component 
$\theta^3_\mu(s)$,
the abelian field strength $\theta^{\rm Abel}_{\mu\nu}$ is defined by
\begin{eqnarray}
\theta_{\mu\nu}^{\rm Abel} &=&
{\rm mod}_{2\pi}(\partial_\mu\theta_\nu^{\rm 3}
-\partial_\nu\theta_\mu^{\rm 3})\\ 
&=& \partial_\mu\theta^{\rm 3}_\nu(s)
-\partial_\nu\theta^{\rm 3}_\mu(s)
+2\pi n_{\mu\nu}(s),
\label{eq:field2}
\end{eqnarray}
where the former part denotes the ordinary two-form 
and $n_{\mu\nu}(s) \in {\bf Z}$
corresponds to the Dirac string on the lattice \cite{DeGrand}.
\noindent 
In the lattice formalism, the photon part
$\theta^{Ph}_\mu(s)$ and the monopole part
$\theta^{Mo}_\mu(s)$ are
obtained from $\theta^{\rm Abel}_{\mu\nu}(s)$ 
and $2\pi n_{\mu\nu}(s)$, respectively
\begin{eqnarray}
\theta^{Ph}_{\mu}(s)&=&
-\{\Box^{-1}\partial_{\nu}\theta^{\rm Abel}_{\mu\nu}\}(s) 
\label{eq:L-def-ph}\\
\theta^{Mo}_{\mu}(s)&=&-2\pi \{\Box^{-1}\partial_{\nu}n_{\mu\nu}\}(s),
\label{eq:L-def-mo}
\end{eqnarray}
using 
the inverse d'Alembertian $\Box^{-1}$ on the lattice 
[22-28]:
The diagonal gluon component $\theta^{\rm 3}_\mu(s)$ is 
found to be decomposed as 
\begin{equation}
\theta^{\rm 3}_\mu(s)=\theta^{Ph}_\mu(s)+\theta^{Mo}_{\mu}(s)
\label{eq:decom2}
\end{equation}
in the Landau gauge, $\partial_{\mu}\theta^{\rm 3}_{\mu}(s)=0$. 

The field strengths, $\theta^{Ph}_{\mu\nu}$ in the photon part
and $\theta^{Mo}_{\mu\nu}$ in the monopole part
are  given as
\begin{eqnarray}
\theta^{Ph}_{\mu\nu} =
{\rm mod_{2\pi}}(\partial_\mu\theta^{Ph}_\nu-\partial_\nu\theta^{Ph}_\mu)\\
\theta^{Mo}_{\mu\nu} =
{\rm mod_{2\pi}}(\partial_\mu\theta^{Mo}_\nu-\partial_\nu\theta^{Mo}_\mu)
\label{eq:field-m}
\end{eqnarray}
on the lattice.
In the continuum limit $a \to 0$,
these field strengths becomes as
\begin{eqnarray}
\theta^{Ph}_{\mu\nu} &\to& \frac{1}{2}a^2eF^{\rm Ph}\\
\theta^{Mo}_{\mu\nu} &\to& \frac{1}{2}a^2eF^{\rm Mo}.
\end{eqnarray}
Using the dual field strength $^*\!\theta^{Mo}_{\mu\nu}\equiv
\frac12 \varepsilon_{\mu\nu\rho\sigma} \theta^{Mo}_{\rho\sigma}$,
the dual Wilson loop is expressed as 
\begin{equation}
W_{D}(C) =
\frac{1}{2}{\rm tr}\exp \left(
i\int_S ~^*\!\theta^{Mo}_{\mu\nu}dS_{\mu\nu} \right)
\label{eq:d-w-lat}
\end{equation} 
using the Stokes theorem
on the lattice.

To summarize, we show the procedure on the derivation of the dual
Wilson loop from lattice QCD as follows:

\begin{enumerate}
\item
We generate the SU(2) gauge configurations $\{U_{\mu}(s)\}_i$
using the Monte Carlo method for the lattice QCD.

\item
We carry out the gauge transformation,
$U_\mu(s) \to U^{\rm MA}_\mu(s)$,
so as to satisfy the 
MA gauge fixing condition with the minimization of
$R_{\rm }$.

\item
The SU(2) link-variable $U_{\mu}(s)$ is factorized as
$U_{\mu}(s)=M_{\mu}(s)u_{\mu}(s)$,
and the abelian projection is performed by the  replacement of  
$U_\mu(s) \in$ SU(2) by 
the abelian link-variable $u_{\mu}(s)=\exp\{ i\theta^{3}_\mu(s)\tau^{3} \}
\in$ U$(1)_3$.

\item
The two-form of the diagonal gluon component is decomposed as
$\partial_\mu\theta^{\rm 3}_\nu(s)
-\partial_\nu\theta^{\rm 3}_\mu(s)
= \theta^{\rm Abel}_{\mu\nu}(s)+2\pi n_{\mu\nu}(s)$
with the abelian field strength $\theta^{\rm Abel}_{\mu\nu}
\in (-\pi,\pi]$ 
and
the Dirac string $2\pi n_{\mu\nu} \in 2\pi{\bf Z}$.

\item
The abelian gauge field $\theta^{\rm 3}_{\mu}$ is 
decomposed as $\theta_{\mu}^{\rm 3}=\theta_{\mu}^{Mo}+\theta_{\mu}^{Ph}$
with the  photon part $\theta_{\mu}^{Ph}$ and the monopole part 
$\theta_{\mu}^{Mo}$, which are obtained from $\theta^{\rm Abel}_{\mu\nu}$
and $n_{\mu\nu}$ by way of Eqs.(\ref{eq:L-def-ph}) and (\ref{eq:L-def-mo})
using the inverse d'Alembert operator 
$\Box^{-1}$ in the Landau
gauge.

\item
Using the field strength $\theta_{\mu\nu}^{Mo}$ in the monopole part, 
the dual Wilson loop 
$\langle W_D(C) \rangle_{\rm MA}$
and the inter-monopole potential $V_M(r)$ are calculated with
Eqs.(\ref{eq:d-w-lat}) and (\ref{eq:imp}).
Then, the effective dual-gluon mass $m_B$ is estimated
from the inter-monopole potential $V_M(r)$.
\end{enumerate}

\noindent
Thus, we investigate the dual Higgs mechanism in QCD 
by extracting the dual Wilson loop 
$\langle W_D(C) \rangle_{\rm MA}$
and the inter-monopole potential $V_M(r)$ from the gauge configurations
obtained in
the lattice QCD simulations.
\section{Lattice QCD Results for Inter-Monopole Potential and Dual Gluon Mass}

In this section, we show the numerical result of 
the lattice QCD simulation.
For the study of the dual Higgs mechanism in QCD,
we calculate the dual Wilson loop $W_D(R,T)$ and 
the inter-monopole potential $V_M(r)$ in the monopole part
(the monopole-current system) in QCD
in the MA gauge
using the SU(2) lattice with $20^4$ and $\beta=2.2 \sim 2.3$.
All measurements are performed at every 100 sweeps
after a thermalization of 5000 sweeps using 
the heat-bath algorithm.
The physical unit or the lattice spacing $a$
is determined so as to reproduce the string tension
$\sigma = 1$ GeV/fm for each $\beta$,
{\it e.g.} $a=0.199$ fm for $\beta=2.3$.
We prepare 100 samples of gauge configurations.
These simulations have been performed using the super-computer SX-4 at 
Osaka University.


The dual Higgs mechanism is characterized by 
the effective-mass acquirement of the dual gluon
$B_\mu$, which is brought by monopole condensation.
To examine the effective mass of $B_\mu$,
we calculate the inter-monopole potential $V_M(r)$
from the dual Wilson loop $\langle W_{D}(R,T) \rangle_{\rm MA}$
obtained in the lattice QCD.
As shown in Fig.1, 
the dual Wilson loop $\langle W_{D}(R,T) \rangle_{\rm MA}$
seems to obey the perimeter law rather than 
the area law
for large loops with  $I,J \geq 3$.
Since the dual Wilson loop $\langle W_{D}(R,T) \rangle_{\rm MA}$
 satisfies the perimeter law as
\begin{equation}
\ln \langle W_{D}(R \times T) \rangle_{\rm MA} \simeq -2(R+T)\cdot \alpha
\label{eq:per-dw}
\end{equation}
for large $R$ and $T$, the inter-monopole potential becomes constant $2 \alpha$
in the infinite limit of $T$,
\begin{equation}
V_{M}(R) \to \lim_{T \to \infty} \frac{2 \alpha}{T}R + 2\alpha = 2 \alpha.
\label{eq:per-v}
\end{equation}
In the actual lattice QCD calculation,
however, we have to take a finite length of $T$, and 
hence
the linear part $(2\alpha /T)R$ slightly remains as a 
lattice artifact [28].
Therefore,
we have to subtract this lattice artifact $(2\alpha /T)R$
for evaluating of the inter-monopole potential $V_M(R)$
from $\langle W_D(R,T) \rangle$ in the lattice QCD simulation. 
Here, $\alpha$ can be estimated from the slope of 
the dual Wilson loop $\ln \langle W_{D}(R \times T) \rangle_{\rm MA}$
for large $R$ and $T$ for each lattices.
We obtain $\alpha=0.141 \pm 0.025$GeV 
from the dual Wilson loop in Fig.1(b) with $\beta=2.3$.

After the subtraction of the lattice artifact $(2\alpha /T)R$,
we consider the slope of the inter-monopole potential $V_M(r)$
 in the lattice QCD. As shown in Fig.2,
the inter-monopole potential $V_M(r)$ is short-ranged and flat
in comparison with the linear inter-quark potential
with string tension $\sigma=1$GeV/fm.
Now, we try to apply the Yukawa potential $V_Y(r)$,
\begin{equation}
V_{Y}(r) = -\frac{(e/2)^2}{4\pi}\frac{e^{-m_Br}}{r}
\label{eq:y-p}
\end{equation}
to the inter-monopole potential $V_M(r)$. As shown in Fig.3, 
the inter-monopole potential
can be fitted  by the Yukawa potential $V_Y(r)$
in the long distance region, and we evaluate 
the dual gluon mass as $m_B \simeq 0.5{\rm GeV}$.

Finally, we consider the possibility of the monopole size effect,
because the QCD monopole is 
expected to be a soliton like object composed of gluons.
In fact, from the recent lattice QCD study,
the QCD monopole includes large off-diagonal gluon
components near its center even in the MA gauge 
[28,30,32],
and the off-diagonal gluon richness would provide 
the ``effective size'' of the QCD monopole
similar to the 't Hooft-Polyakov monopole 
[1-4]
We introduce the effective size $R_{\rm M}$ of the QCD-monopole, and
assume the Gaussian-type distribution of the magnetic charge
around its center,
\begin{equation}
\rho({\bf x};R_{\rm M}) = \frac{1}{(\sqrt{\pi}R_{\rm M})^3}\exp({\frac{-\vert{\bf 
x}\vert^2}{R_{\rm M}^2}}).
\label{eq:mc-den}
\end{equation}
Since the monopole part is an abelian system,
simple superposition on $B_\mu$
is applicable like the Maxwell equation.
Therefore, the inter-monopole potential
with the effective monopole size $R_{\rm M}$
is expected to be
\begin{equation}
V({\bf x};R_{\rm M}) = 
-\frac{(e/2)^2}{4\pi}\int d^3x_1 \int d^3x_2 \rho({\bf x}_1;R_{\rm M})
\rho({\bf x}_2;R_{\rm M})
\frac{\exp({-m_B\vert{\bf x-x_1+x_2}\vert})}{
\vert{\bf x-x_1+x_2}\vert},
\label{eq:y-t-p}
\end{equation}
\noindent
or equivalently
\begin{eqnarray}
V(r;R_{\rm M}) &=& -\frac{(e/2)^2}{\pi^2 R_{\rm M}^6} \int^\infty_0 dr_1 \int^\infty_0
 dr_2 e^{-(r_1^2+r_2^2)/R_{\rm M}^2}
\int^\pi_0 d\theta_1 \int^\pi_0 d\theta_2
\sin\theta_1\sin\theta_2 \nonumber \\
\times&& \!\!\!\!\!\!\!\!\!
\frac{\exp[-m_B\sqrt{ \{ \sqrt{(r-r_2\cos\theta_2)^2+(r_2\sin\theta_2)^2}
-r_1\cos\theta_1 \}^2+(r_1\sin\theta_1)^2}]} 
{
\sqrt{ \{ \sqrt{(r-r_2\cos\theta_2)^2+(r_2\sin\theta_2)^2}
-r_1\cos\theta_1 \}^2+(r_1\sin\theta_1)^2}}, 
\label{eq:y-t-p2}
\end{eqnarray}
where $r \equiv |{\bf x}-{\bf y}|$
 is the distance between the two monopole centers.
We apply the Yukawa-type potential $V(r;R_{\rm M})$
to the inter-monopole potential $V_M(r)$ 
in Fig.3. 
The potential $V(r;R_{\rm M})$ with the effective
monopole size $R_{\rm M}=0.21$fm seems to fit
the lattice data of $V_M(r)$ in the whole region of the 
distance $r$.

Thus, we estimate the dual gluon mass $m_B \simeq 0.5$GeV
and the effective monopole size $R_{\rm M} \simeq 0.2$fm
by evaluating the inter-monopole potential $V_M(r)$
from the dual Wilson loop $W_D(R,T)$ in the monopole part in the MA gauge.
In the long-distance region,
we find the effective-mass acquirement of the dual gluon $B_\mu$,
which is essential for the dual Higgs mechanism
in the dual superconductor scenario.
This result suggests ``infrared monopole condensation''
or monopole condensation in the long-scale description of the QCD vacuum.
The monopole size $R_{\rm M}$ would 
provide a critical scale [30-32] for the nonperturbative QCD
in terms of the dual Higgs theory,
because the QCD-monopole structure such as 
off-diagonal gluons [28,30] should be considered at the shorter
scale than $R_{\rm M}$,
similar to the structure of 
the 't~Hooft-Polyakov monopole.

\section{Summary and Concluding Remarks}
To examine the dual superconductor picture
for the quark confinement mechanism  
in the QCD vacuum, we 
have studied the dual Higgs mechanism in terms of 
the effective-mass acquirement of the dual gluon field
$B_\mu$ using the lattice QCD Monte Carlo simulation.
In the MA gauge, QCD is reduced to the abelian gauge theory
with the color-electric current $j_\mu$ and 
the color-magnetic monopole current $k_\mu$.
The abelian-projected QCD, the diagonal part of QCD, can be separated
into the photon part and
the monopole part
corresponding to the separation of $j_\mu$ and $k_\mu$,
respectively.
Reflecting the abelian dominance and the monopole dominance,
the monopole part carries essence of NP-QCD and then is of interest
in the MA gauge,
so that
we have concentrated ourselves to the monopole part
(the monopole-current system) in QCD.

In order to investigate the dual Higgs mechanism in QCD, 
we have introduced the dual gluon field $B_\mu$
and have studied its features 
in the monopole part in the MA gauge in the lattice QCD.
Owing to the absence of the electric current,
the monopole part resembles the dual version of QED,
and hence this part
is naturally described by the dual gluon field $B_\mu$
without meeting the difficulty on the Dirac-string singularity.
In the dual gauge formalism, the dual Higgs mechanism
is characterized by the acquirement of the effective mass $m_B$
of the dual gluon field $B_\mu$.
Then, 
to evaluate the dual gluon mass,
we have calculated the dual Wilson loop
$\langle W_D(R \times T) \rangle_{\rm MA}$, and have studied the 
inter-monopole potential $V_M(r)$
in the monopole part in the MA gauge
based on the dual gauge formalism by using 
the lattice QCD simulation.

In the lattice QCD,
we have found that 
the dual Wilson loop
obeys the perimeter law for large loops.
Considering the finite-size effect of the dual Wilson loop,
we have investigated the inter-monopole potential $V_M(r)$,
and have found that
$V_M(r)$ is short-ranged and flat
in comparison with the linear inter-quark potential.
Then, we have compared 
the inter-monopole potential $V_M(r)$ with the Yukawa potential and 
have estimated the dual gluon mass as
$m_B \simeq {\rm 0.5GeV}$, 
which is consistent with 
the phenomenological parameter fitting 
in the dual Ginzburg-Landau theory \cite{SST}.
The generation of the dual gluon mass $m_B$ 
in the infrared region
suggests 
the realization of the dual Higgs mechanism 
and monopole condensation
in the long-scale  description of the QCD vacuum.
In this way, we have shown the evidence of ``infrared monopole condensation''
in the lattice QCD in the MA gauge.

To explain the short-range deviation between 
the inter-monopole potential $V_M(r)$
and the Yukawa potential,
we have considered the effective size $R_{\rm M}$ of the monopole,
since the monopole would be a soliton-like object composed of gluons
[28,30-32].
The lattice data of the inter-monopole potential can be well fitted with
the Yukawa-type potential $V(r;R_{\rm M})$ with the effective size
$R_{\rm M} \simeq 0.2$ fm of the monopole.
This monopole size $R_{\rm M} \simeq 0.2$ fm
may provide the critical scale for the dual Higgs theory in QCD,
because the monopole structure relating to off-diagonal gluons
become visible 
[28,30-32]
and the QCD system
cannot be described only with the abelian local field theory
at the shorter scale than $R_M$.

\section*{Acknowledgment}
We would like to thank Professor Hiroshi Toki
for his useful comments and discussions.
One of authors (H.S.) is supported in part by Grant for 
Scientific Research (No.09640359) from the Ministry of Education,
Science and Culture, Japan.
The lattice QCD simulations have been performed on the super-computer 
SX4 at Osaka university.

\baselineskip 1cm

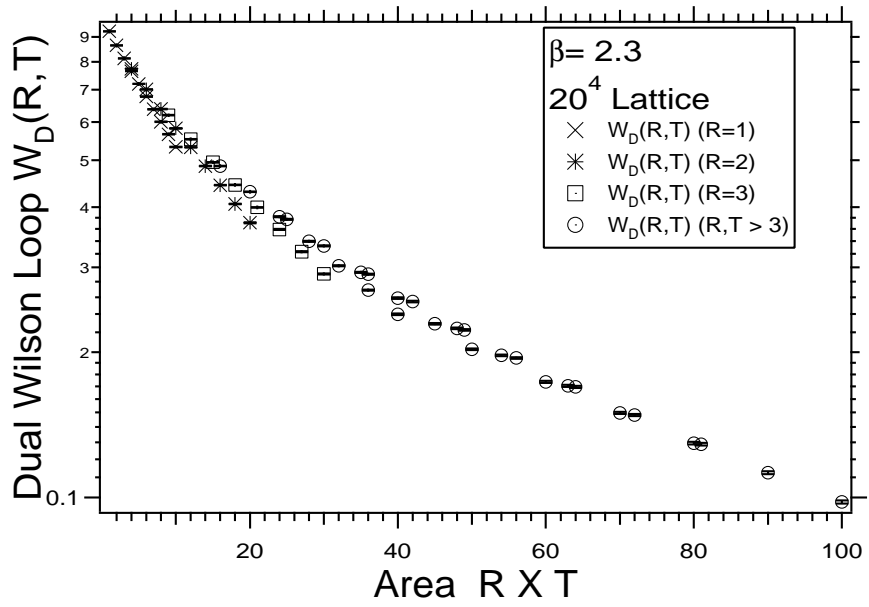
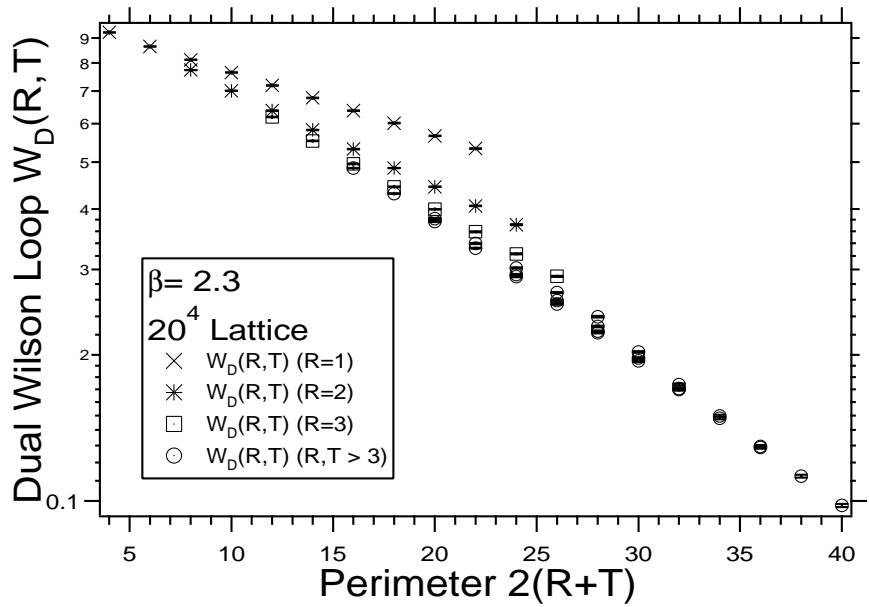
\begin{figure}

\caption{\baselineskip .9cm
(a) The dual Wilson loop $\langle W_D(R,T) \rangle_{\rm MA}$
as the function of its area $R \times T$
(b) $\langle W_D(R,T) \rangle_{\rm MA}$ as the function of 
its perimeter $L \equiv 2(R+T)$ in the monopole part
in the MA gauge in the SU(2) lattice QCD
with $20^4$ and $\beta=2.3$.
For large loops as $R,T \geq 3$,
$\langle W_D(R,T) \rangle_{\rm MA}$
seems to obey the perimeter law rather than
the area law.
\label{fig:dw}
} 
\end{figure}
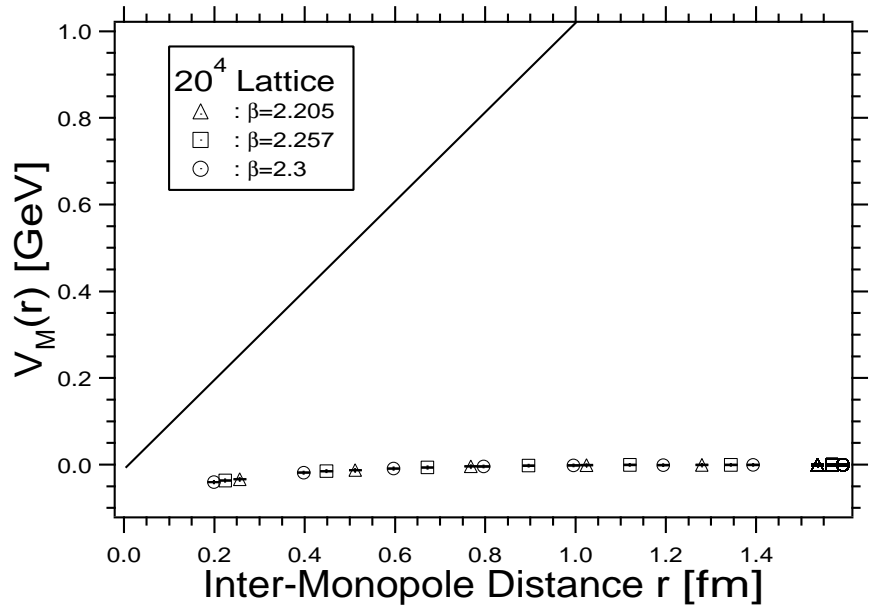
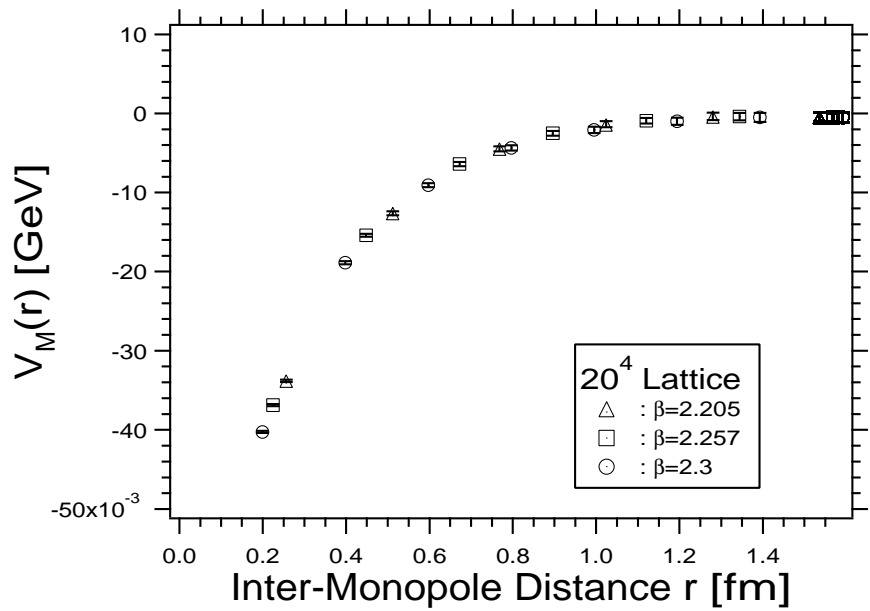
\begin{figure}

\baselineskip 1cm

\caption{\baselineskip .9cm
(a) The inter-monopole potential
$V_M(r)$ as the function of the inter-monopole 
distance $r$ in the monopole part in the MA gauge in the SU(2) lattice QCD
with $20^4$ lattice and $\beta=2.2 \sim 2.3$.
For comparison, we plot also the linear part of
the inter-quark potential $V_q^{\rm linear}(r)=\sigma r$
with $\sigma=1.0$GeV/fm by the straight line. (b) The detail
of the lattice QCD data for the inter-monopole potential $V_M(r)$.
\label{fig:poten} 
}
\end{figure}

\baselineskip 1cm

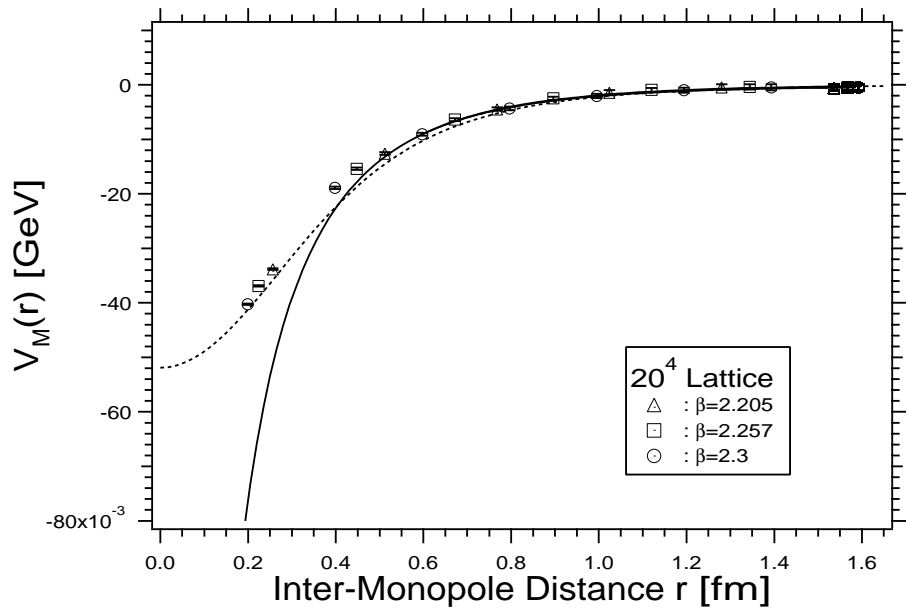
\begin{figure}
\caption{\baselineskip .9cm
The analysis for the shape of the inter-monopole potential $V_M(r)$ 
in the SU(2) lattice QCD with $20^4$ and $\beta=2.2 \sim 2.3$.
The solid curve denotes the simple Yukawa potential $V_Y(r)$
with the dual gluon mass $m_B=0.5$ GeV.
The dotted curve denotes the Yukawa-type potential $V(r;R_M)$
including the magnetic-size effect.
The lattice data of the inter-monopole potential $V_M(r)$
seem to be fitted by $V(r;R_{\rm M})$ 
with the effective monopole size $R_{\rm M} \simeq {\rm 0.2fm}$
in the whole region of r. 
\label{fig:poten2}
}
\end{figure}

~\\
\newpage
\begin{figure}
\begin{center}
\epsfig{figure=dw-a.EPSF,height=8cm}
\centerline{(a)}
\\
\vspace{2.0cm}
\epsfig{figure=dw-p.EPSF,height=8cm}
\centerline{(b)}
\\
\vspace{1.0cm}
{\Huge Figure \ref{fig:dw}}
\end{center}
\end{figure}
\newpage

\begin{figure}
\begin{center}
\epsfig{figure=imp-quark.EPSF,height=8cm}
\centerline{(a)}
\\
\vspace{2.0cm}
\epsfig{figure=imp.EPSF,height=8cm}
\centerline{(b)}
\\
\vspace{1.0cm}
{\Huge Figure \ref{fig:poten}}
\end{center}
\end{figure}
\newpage

\begin{figure}
\begin{center}
\epsfig{figure=imp-r02.EPSF,height=8cm}
\\
\vspace{1.0cm}
{\Huge Figure \ref{fig:poten2}}
\end{center}
\end{figure}


\begin{thebibliography}{99}
\bibitem{Pokorski} S. Pokorski,
                ``Gauge Field Theories''
                (Cambridge University Press, 1985) 1.
\bibitem{Cheng}
                For instance, T.~P.~Cheng and L.~F.~Li,
                ``Gauge Theory of Elementary Particle Physics''
                (Clarendon press, Oxford, 1984) 1.
\bibitem{Itzykson}For instance, C.~Itzykson and J.-B.~Zuber,
                ``Quantum Field Theory'' (McGraw-Hill, New York, 1985) 1.

     \bibitem{Huang}
K. Huang, ``Quarks, Leptons and Gauge Fields'', (World Scientific,
Singapore, 1991) 1.

\bibitem{Aitchison} I. J. R. Aitchison and A. J. G. Hey,
                ``Gauge Theories in Particle Physics''
                (Adam Hilger, Bristol and Philadelphia, 1992) 1.


      \bibitem{Greiner}
 W.~Greiner and A.~Schafer, ``Quantum Chromodynamics'', (Springer,1994) 1.

      \bibitem{Rothe}
H.~J.~Rothe, ``Lattice Gauge Theories'', (World Scientific, 1992) { Sov. Phys. }JETP {\bf 5}, 1174 (1957).1. 


\bibitem{Nambu} Y. Nambu, \Journal{\PRD}{10}{1974}{4262}.

\bibitem{tHooft75}      G. 't Hooft,
                \MEETtmp{in}
                {High Energy Physics}
                {edited by A. Zichichi}
                {Editorice Compositori, Bologna, 1975}.

\bibitem{Mandelstam}S. Mandelstam,  \Journal{\PRC}{23}{1976}{245}.


\bibitem{Abrikosov} A. A. Abrikosov, 
                ``Fundamentals of the Theory of Metals''
                (Adam Hilger, 1988) 1.

\bibitem{Haymaker} R. W. Haymaker, V. Singh, Y.-C. Peng, and J. Wosiek, 
		\Journal{\PRD}{53}{1996}{389}.
V. Singh, D. A. Browne, R. W. Haymaker, 
		\Journal{\PLB}{306}{1993}{115}.


\bibitem{SST}H. Suganuma, S. Sasaki, and H. Toki,
                \Journal{\NPB}{435}{1995}{207}.\\
H.~Ichie, H.~Suganuma and H.~Toki, Phys. Rev. {\bf D52}~2994~(1995);
{\bf D54}~3382~(1998). \\
S.~Umisedo, H.~Suganuma and H.~Toki,   
Phys. Rev.{\bf D57} 1605~(1998). 

\bibitem{tHooft81}G. 't Hooft,  \Journal{\NPB}{190}{1981}{455}.

\bibitem{Ezawa} Z. F. Ezawa and A. Iwazaki,
                {Phys. Rev.} D {\bf25}, 2681 {(1982)} ;
                {\bf 26}, {631} {(1982)}.


\bibitem{Kronfeld}A. S. Kronfeld, G. Schierholz, and U.-J. Wiese,
                \Journal{\NPB}{293}{1987}{461}.

\bibitem{Brandstater}    F. Brandstater, U.-J. Wiese, and G. Schierholz,
                \Journal{\PLB}{272}{1991}{319}.


\bibitem{Yotsuyanagi} 
T. Suzuki and I. Yotsuyanagi, \Journal{\PRD}{42}{1990}{4257}.

\bibitem{Hioki91}  S. Hioki, S. Kitahara, S. Kiura, Y. Matsubara, O. Miyamura,
                S. Ohno, and T. Suzuki, \Journal{\PLB}{272}{1991}{326}.

     \bibitem{Miyamura}
O.~Miyamura, Phys.~Lett.~{\bf B353} 91 (1995);
Nucl.~Phys.~{\bf B} (Proc. Suppl.) {\bf 42} 538 (1995). \\
O.~Miyamura and S.~Origuchi,
``Color Confinement and Hadrons (Confinement '95)",
edited by H.~Toki, Y.~Mizuno, H~Suganuma, T.~Suzuki and O.~Miyamura, 
(World Scientific, 1995) 65.

\bibitem{Woloshyn}R.~M.~Woloshyn, \Journal{\PRD}{51}{1995}{6411}.


     \bibitem{DeGrand}
 T. DeGrand and D. Toussaint, Phys. Rev. {\bf D22} 2478 (1980).

 \bibitem{Stack_Wensley} J. D. Stack, R. J. Wensley and S. D. Heiman,
                \Journal \PRD{50}{1994}{3399}.

      \bibitem{Bali}
G.~S.~Bali, V.~Bornyakov,  M.~Muller-Preussker and K.~Schilling,
Phys. Rev. {\bf D54}~2863 (1996).


	\bibitem{Polikarpov} 
A.~Di~Giacomo,  Nucl.~Phys.~{\bf B} (Proc. Suppl.) {\bf 47} 136 (1996)
and references therein.    
M.~I.~Polikarpov, Nucl.~Phys.~{\bf B} (Proc. Suppl.) {\bf 53} 134 (1997)
and references therein.

 \bibitem{STSM} H. Suganuma, A~Tanaka, S~Sasaki, O~Miyamura,
Nucl.~Phys.~{\bf B} (Proc. Suppl.) {\bf 47} 302 (1996).\\
H. Suganuma, M. Fukushima, H. Ichie, and A. Tanaka,
                \Journal \NPBPS{65}{1998}{29}.

      \bibitem{INNOCOM_Tanaka}
A.~Tanaka and H.~Suganuma, Proc. of Int. Symp. 
on ``Innovative Computational Methods in Nuclear Many-Body Problems'', Osaka, Nov. 1997,
(World Scientific) 281.  

\bibitem{YKIS_Suganuma}H. Suganuma, H. Ichie, A. Tanaka, and K. Amemiya,
                \Journal{\PTPS}{131}{1998}{559}.
      \bibitem{Blagojevic}
 M.~Blagojevic and P.~Senjanovic, Nucl.~Phys.~{\bf B161} 112 (1979).

	\bibitem{HI_HS} 
	H.~Ichie and H.~Suganuma,
                preprint, hep-lat/9808054.

      \bibitem{QULEN_Ichie}
H.~Ichie, H.~Suganuma and A.~Tanaka,
Nucl.~Phys.~{\bf A629}, A629, 82c (1998).  



       \bibitem{INNOCOM_Ichie}
 H.~Ichie and H.~Suganuma, Proc. of Int. Symp. 
 on ``Innovative
 Computational Methods in Nuclear Many-Body Problems'', Osaka, Nov. 1997, 
 (World Scientific) 278,  hep-lat/9802032.
 



 


     

\end{thebibliography}
\end{document}